\documentclass[10pt]{article}

\usepackage{amsmath}
\usepackage{amssymb}

\usepackage{graphicx}
\usepackage{subfigure}
\usepackage{cite}

\usepackage{color}

\usepackage[labelfont=bf,labelsep=period,justification=raggedright]{caption}

\makeatletter
\renewcommand{\@biblabel}[1]{\quad#1.}
\makeatother

\date{}

\begin{document}

\begin{flushleft}
{\Large \textbf{Information Filtering via Balanced Diffusion on Bipartite Networks} }

\bigskip Da-Cheng NIE, Ya-Hui AN, Qiang DONG$^\ast$, Yan FU, Tao ZHOU 

\bigskip Web Sciences Center, School of Computer Science \& Engineering,
University of Electronic Science and Technology of China, Chengdu
611731, China

\bigskip$\ast$ Corresponding Author. E-mail address: dongq@uestc.edu.cn (Q.
Dong).
\end{flushleft}

\section*{Abstract}

Recent decade has witnessed the increasing popularity of recommender
systems, which help users acquire relevant commodities and services
from overwhelming resources on Internet. Some simple physical diffusion
processes have been used to design effective recommendation
algorithms for user-object bipartite networks, typically mass diffusion (MD) and heat conduction (HC) algorithms which have different advantages respectively on accuracy and diversity. In
this paper, we investigate the effect of weight assignment in the
hybrid of MD and HC, and find that a new hybrid algorithm of
MD and HC with balanced weights will achieve the optimal
recommendation results, we name it balanced diffusion (BD) algorithm. Numerical
experiments on three benchmark data sets, $MovieLens$, $Netflix$ and $RateYourMusic (RYM)$, show that the performance of BD algorithm outperforms
the existing diffusion-based methods on the three important recommendation
metrics, accuracy, diversity and novelty. Specifically, it can not only provide accurately recommendation results, but also yield higher diversity and novelty in recommendations by accurately recommending unpopular objects.

\section*{Author Summary}
Conceived and designed the experiments: DCN QD TZ. Performed the
experiments: DCN YHA. Analyzed the data: DCN YHA QD YF TZ. Contributed
reagents/materials/analysis tools: DCN YHA YF TZ. Wrote the paper:
DCN YHA QD TZ.
\section*{Introduction}

Web 2.0 and its applications have achieved significant developments
in the past few years, which bring us more convenience as well as
overwhelm us with the information ocean on Internet
\cite{froomkin1995flood}. This is the so-called \emph{Information
Overload} problem \cite{edmunds2000problem}. Nowadays, online
shopping becomes more and more popular in our daily life. For
instance, there are millions of books (e-books) on
\emph{Amazon.com}, and the value of transactions of
\emph{Taobao.com} exceeded 35 billion Chinese Yuan (about 6 billion
US dollars) on the shopping festival of Nov 11, 2013 \cite{liu2013china}. In this
case, we find that it is very difficult to choose the relevant ones
from countless candidates on these e-commerce websites, and thus an
automatic way that can help us to make right decision under the
information overload becomes a significant issue for both academic
and industrial communities.

The emergence of search engines partially alleviates this dilemma; a
user inputs the keywords and then search engines return the
results accordingly. However, search engines always return the same
results for different users if they key in the same words. When
users resort to the search engines, they have already known what
they want, and can easily find keywords to describe it. But in the
most occasions, users do not know what they really want, or it is
hard to find appropriate words to describe it. Therefore,
recommender systems have been designed to solve this problem. We can
see that in recent years, recommender systems have greatly promoted
the development of E-business, and vice versa \cite{dias2008value}.

Collaborative filtering
(CF)\cite{adomavicius2005toward,schafer2007collaborative,herlocker1999algorithmic,sarwar2001item,deshpande2004item,resnick1994grouplens,breese1998empirical,linden2003amazon}
is the most frequently used technology in the recommender systems,
which makes use of the collecting history to predict the potential
objects of interest to the target user, including user-based CF
\cite{resnick1994grouplens}  and item-based CF
\cite{sarwar2001item,linden2003amazon}. However, the classical CF
methods only take user/item similarity into account, which will make the recommendation results more and more similar among users. In
the meanwhile, CF algorithms cannot deal with the cold start problem
\cite{schein2002methods}, i.e., when a new user or object is added
to the system, it is difficult to obtain recommendations or to be
recommended. Therefore, the content-based \cite{pazzani2007content}
methods have been proposed to solve this problem, which use the
profiles of users to generate recommendation results, but the user
profiles are usually difficult to acquire due to the constraint of
information retrieval techniques. Generally speaking, the CF methods
and content-based methods will generate similar recommendation
results with poor diversity and novelty.

To address this problem, many personalized recommendation algorithms
have been proposed, including trust-aware methods
\cite{burke2007hybrid,palmisano2008using}, social-impact
methods \cite{nie2013social} and tag-aware methods
\cite{zhang2011tag}. Recently, some physical diffusion processes, have been used to design many
effective diffusion-based recommendation algorithms for user-object
bipartite networks, such as the mass diffusion (MD) algorithm and
heat conduction (HC) algorithm
\cite{zhou2010solving,zhou2007bipartite,zhang2007heat}.
The MD algorithm is in fact a random walk process on the
bipartite network \cite{zhou2007bipartite,huang2004applying}, which
achieve high accuracy but low diversity. HC is another process
similar to MD on the bipartite network, which has high diversity
but low accuracy. Ideally, a good recommendation algorithm should
exhibit both of high accuracy and high diversity.

In \cite{zhou2010solving}, Zhou et al. proposed a hybrid method to
nonlinearly combine the MD and HC algorithms (HHP for short),
which solves the apparent diversity-accuracy dilemma of recommender
systems. Liu et al. \cite{liu2011information} proposed a biased heat
spreading (BHC for short) algorithm, which simultaneously enhances
the accuracy and diversity. L\"{u} et al. \cite{lu2011information}
proposed a hybrid algorithm of ProbS and HeatS with preference on
mass diffusion, named PD algorithm. All of the above mentioned
algorithms derived from MD and HC demonstrate good accuracy
and diversity. However, how to adjust the weights of MD and HC
to get the optimal accuracy and diversity remains to be an open
problem.

In this paper, we investigate the effect of weight assignment in the
hybrid of MD and HC, and find that a new hybrid algorithm of
MD and HC with balanced weights will achieve the optimal
recommendation results, we name it balanced diffusion (BD) algorithm. Numerical
experiments on three benchmark data sets, $MovieLens$, $Netflix$ and
$RateYourMusic$ ($RYM$), show that the performance of BHP algorithm
outperforms the existing diffusion-based methods on the three important
recommendation metrics, accuracy, diversity and novelty.


\section*{Results}
A recommender system can be represented by a bipartite
network $G(U,O,E)$, where $U = \{u_1,u_2,\cdots,u_m\}$, $O =
\{o_1,o_2, \cdots ,o_n\}$, and $E = \{e_1,e_2, \cdots ,e_q \}$ represent the $m$ users, $n$ objects, and $q$ links between the $m$ users and $n$ objects, respectively. The system could be fully described by an adjacency matrix $A = \{a_{l\alpha}\}_{m,n}$, where $a_{l\alpha} = 1$ if there exists a link $e_i$ between user $l$ and object $\alpha$ and  $a_{l\alpha} = 0$ otherwise.

We assume that a user collects an object because he/she likes it,
then the essential task of a recommender system becomes to generate
a ranking list of the target user's uncollected objects. All the
recommendation algorithms inspired by diffusion-like process work by
initially assigning all the objects a certain amount of resources,
denoted by the vector $\textbf{f}$ (where $f_{\alpha}$ is the
resource of object $o_{\alpha}$), and then reallocating these
resources via the transformation $\textbf{f}' = W\textbf{f}$, where
$W$ is called the resource transfer matrix.

In order to investigate the effect of weight assignment in the
hybrid of MD and HC, we give MD and HC two separate
parameters $a$ and $b$ in the transfer matrix $W$:
\begin{center}
\begin{equation}
\label{equ6} w_{\alpha \beta} =
\frac{1}{k_{o_{\alpha}}^{a}k_{o_{\beta}}^{b}}
\sum_{l=1}^{m}{\frac{a_{l\alpha }a_{l\beta}}{k_{u_l}}},
\end{equation}
\end{center}
when $0\leq a\leq1, b=0$ gives us the BHC algorithm,  $a=1,b=0$
gives us the pure HC algorithm and $a=0, b=1$ gives us the pure
MD algorithm. When $a+b=1,0<a, b<1$ gives us the HHP method, it
seems that the HHP method is the best method to combine MD and
HC today. If $a=b>0$, the Equ.\ref{equ6} can be revised using
only one parameter $\lambda$:
\begin{center}
\begin{equation}
\label{equ7} w_{\alpha \beta} =
\frac{1}{(k_{o_{\alpha}}k_{o_{\beta}})^\lambda}
\sum_{l=1}^{m}{\frac{a_{l\alpha }a_{l\beta}}{k_{u_l}}},
\end{equation}
\end{center}
we call this algorithm Balanced diffusion (BD for short),
where MD and HC obtain the same weights in recommender system.
In this sense, the influence of large degree objets would be
strengthened in the second and last diffusion step if $\lambda<1$
and depressed if $\lambda>1$.

\begin{figure}
  \centering
  \subfigure[]{
    \label{fig:subfig:a} 
    \includegraphics[width=0.31\textwidth]{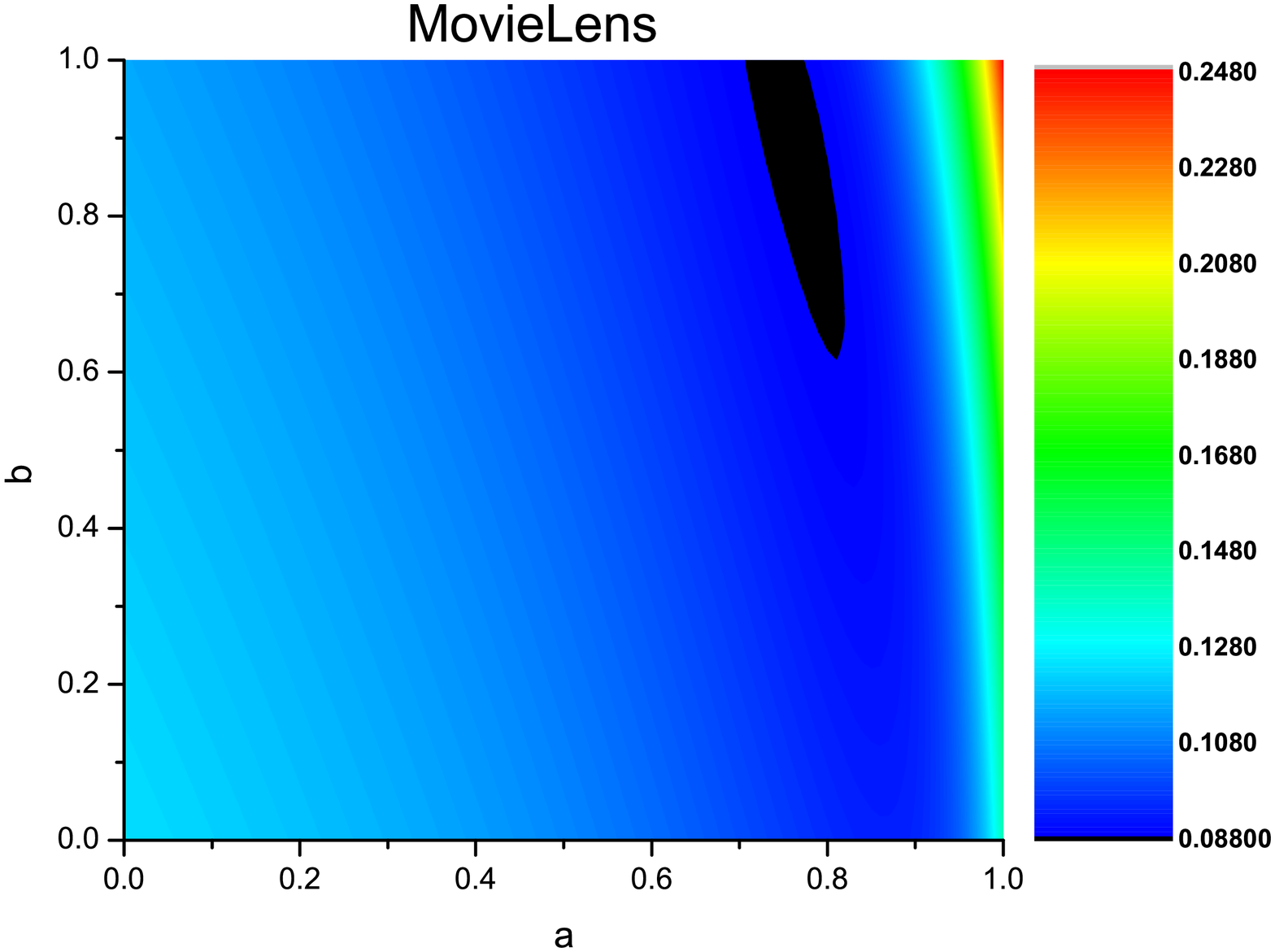}}
  \subfigure[]{
    \label{fig:subfig:b} 
    \includegraphics[width=0.31\textwidth]{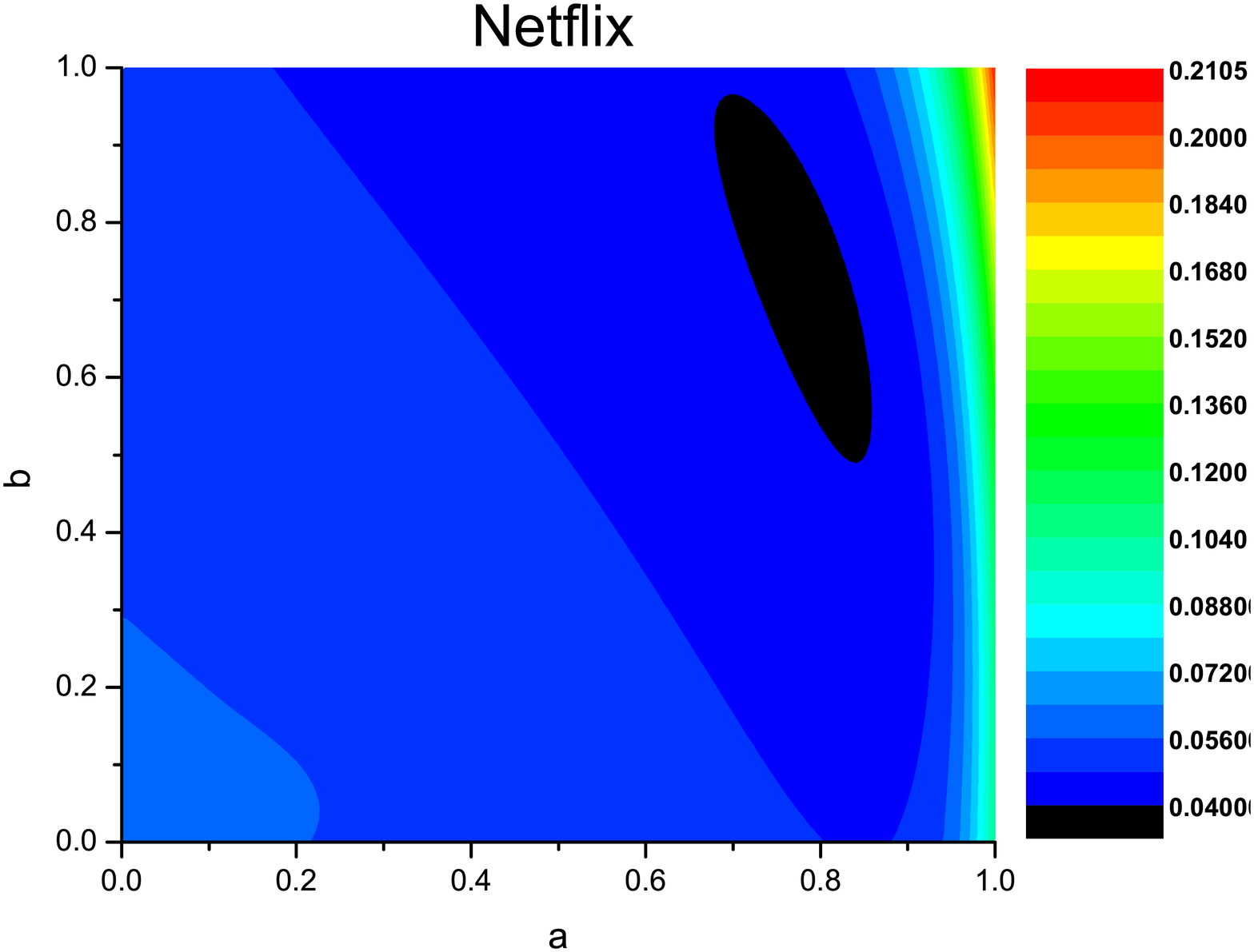}}
  \subfigure[]{
    \label{fig:subfig:c} 
    \includegraphics[width=0.31\textwidth]{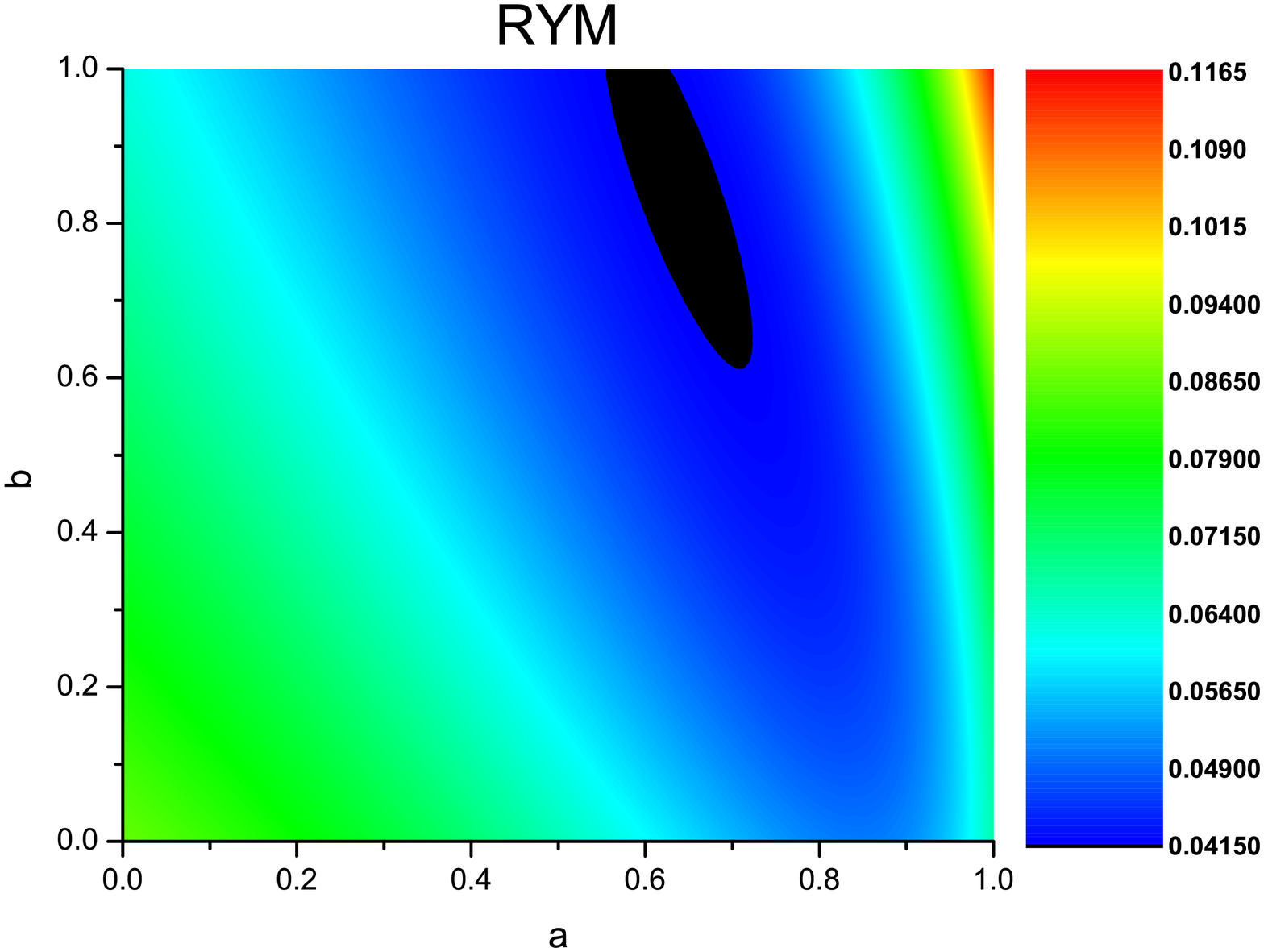}}
  \caption{(Color online) The ranking score on $MovieLens$,
  $Netflix$ and $RYM$ data sets according to Eq.\ref{equ6}}
  \label{fig:subfig} 
\end{figure}

Fig.\ref{fig:subfig:a}, Fig.\ref{fig:subfig:b} and
Fig.\ref{fig:subfig:c} show the ranking score by Equ.\ref{equ6} on
three benchmark data sets, $MovieLens$, $Netflix$ and $RYM$,
respectively. In the figure, different colors represent different
$r$ and the black area means the minimal ranking score values. The
minimal ranking score is 0.087, 0.039 and 0.041 on $MovieLens$,
$Netflix$ and $RYM$, respectively. By comparing the black areas we
find the ranking scores are almost tiny variation. That is to say, in
this area it is difficult to change
the performance of recommender algorithms by adjusting the tunable parameters. In other words, it is not
worth spending too much effort on adjusting the parameters to obtain
the tiny improvement. We find they have a common feature in
the minimal ranking score area, that is, all of them obtain
approximately optimal performance when the two parameters satisfy the condition:
$0<a=b<1$. That is to say,  when MD and HC work together, we need
strength the small-degree objects' influence to obtain the better
performance both at the first diffusion step (i.e., MD) and last diffusion step (i.e., HC). Therefore, we could only use one parameter $\lambda$ to
replace the two in the optimal scenario (i.e., Eq.\ref{equ7}).

\begin{figure}
  \centering
  \subfigure{
    \label{fig1:subfig:a} 
    \includegraphics[width=0.3\textwidth]{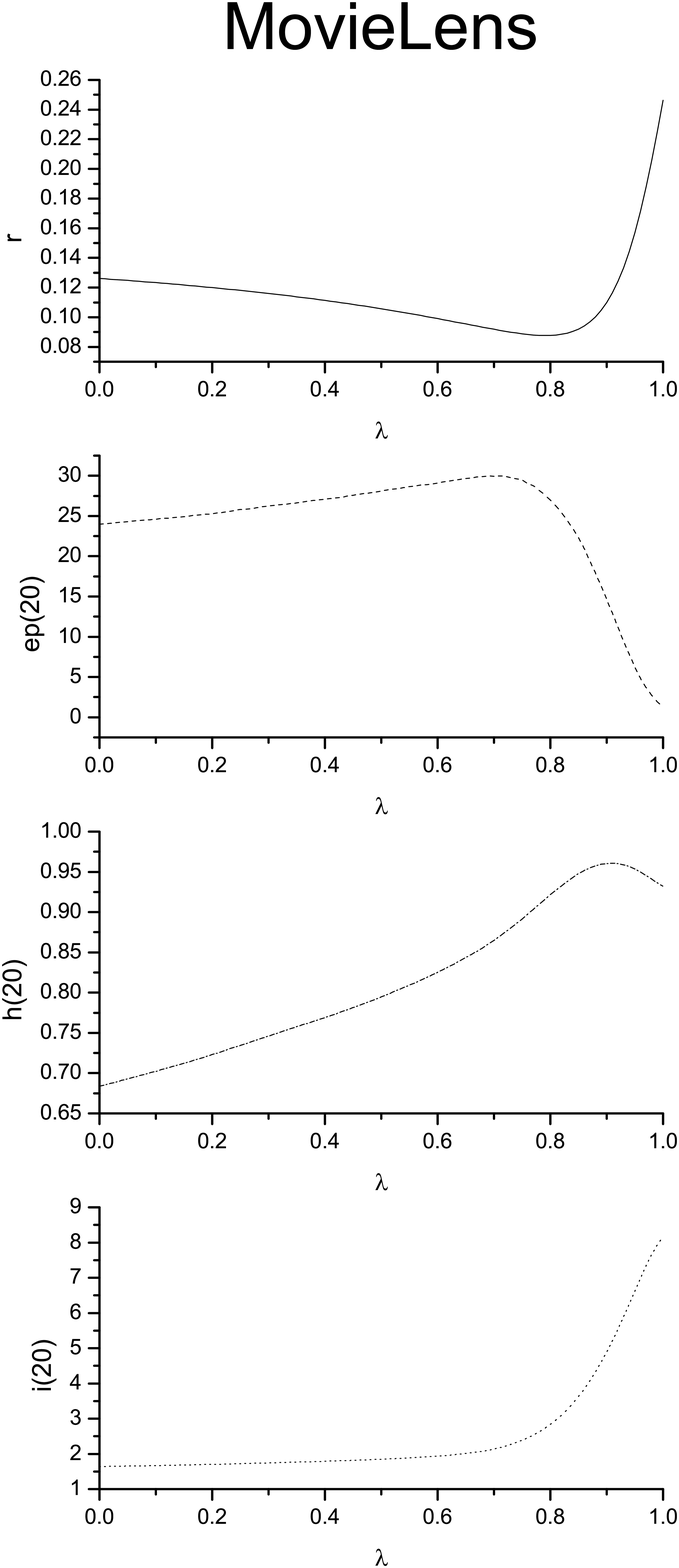}}
  \subfigure{
    \label{fig1:subfig:b} 
    \includegraphics[width=0.3\textwidth]{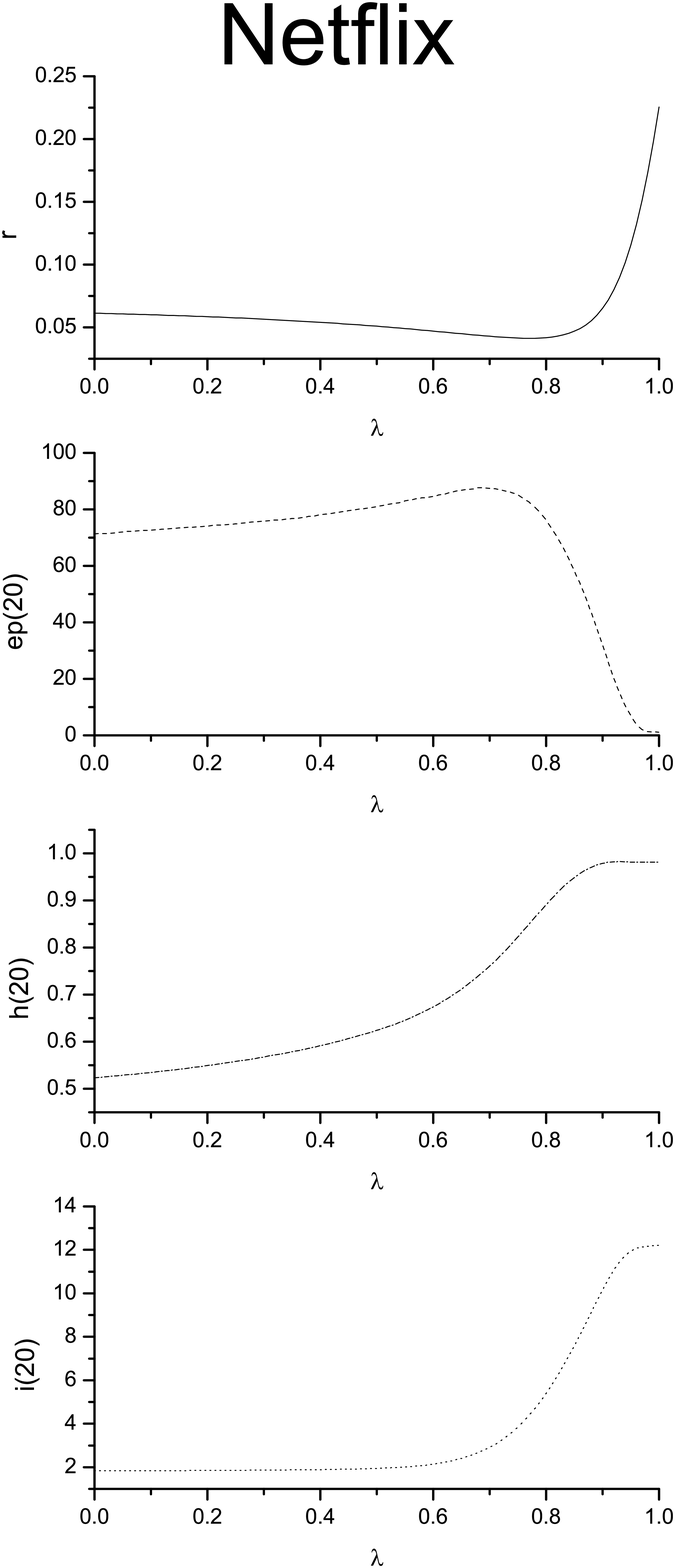}}
  \subfigure{
    \label{fig1:subfig:c} 
    \includegraphics[width=0.3\textwidth]{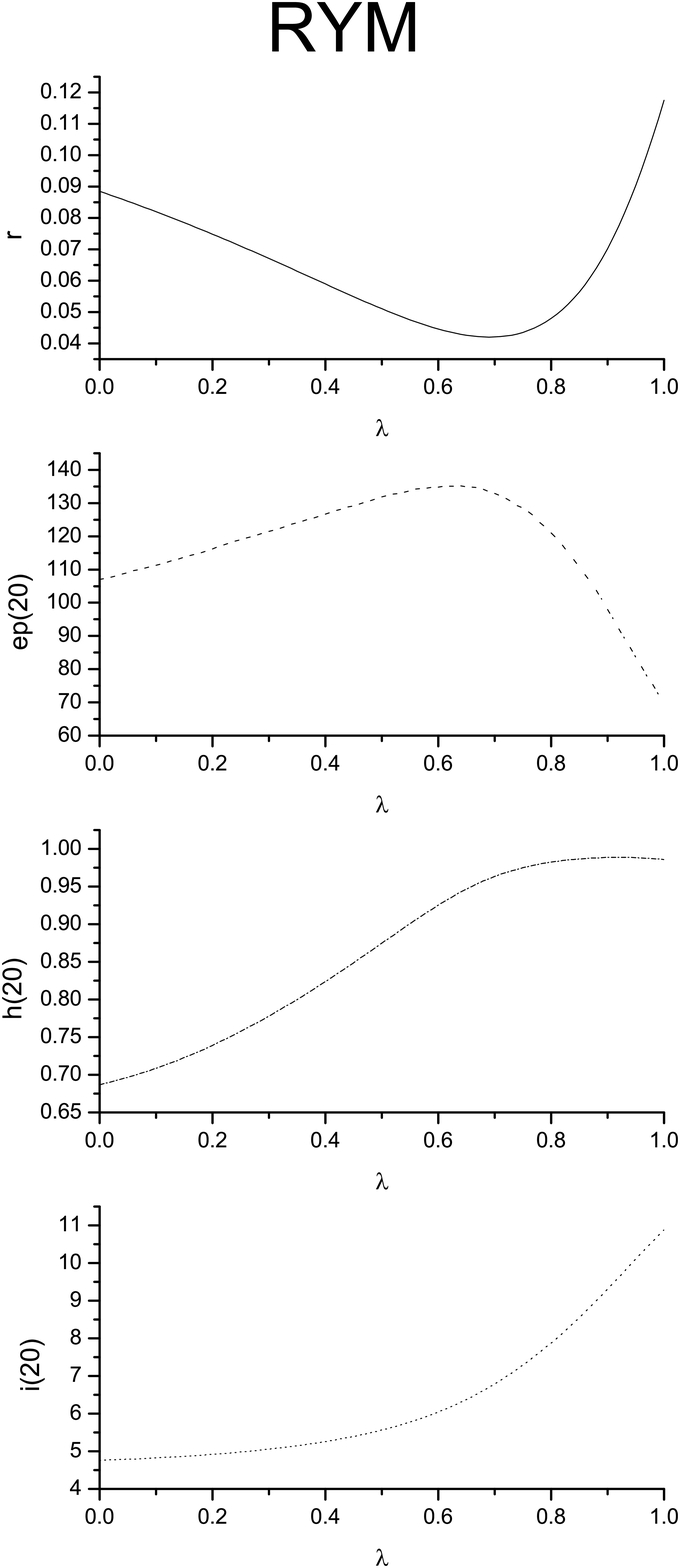}}
  \caption{Performance of the BD algorithm on three different data sets.}
  \label{fig1} 
\end{figure}

Fig.\ref{fig1} shows the performance of BD algorithm on
$MovieLens$, $Netflix$ and $RYM$, respectively. We have known that
the HHP algorithm is a good trade-off of the diversity and accuracy.
From our experiments, we can see that BD is also a good
trade-off of the diversity and accuracy. In other words, our
algorithm improve the diversity and accuracy simultaneously. The
so-called \emph{optimal parameter} depends on the smallest ranking
score. The other three metrics, precision, hamming distance and
novelty's optimal values are obtained at the optimal parameter
point. The optimal parameter of our algorithm is 0.79, 0.77 and 0.69
on $MovieLens$, $Netflix$ and $RYM$ data sets, respectively.

\begin{figure}
  \centering
  \subfigure{
    \label{fig2:subfig:a} 
    \includegraphics[width=3in]{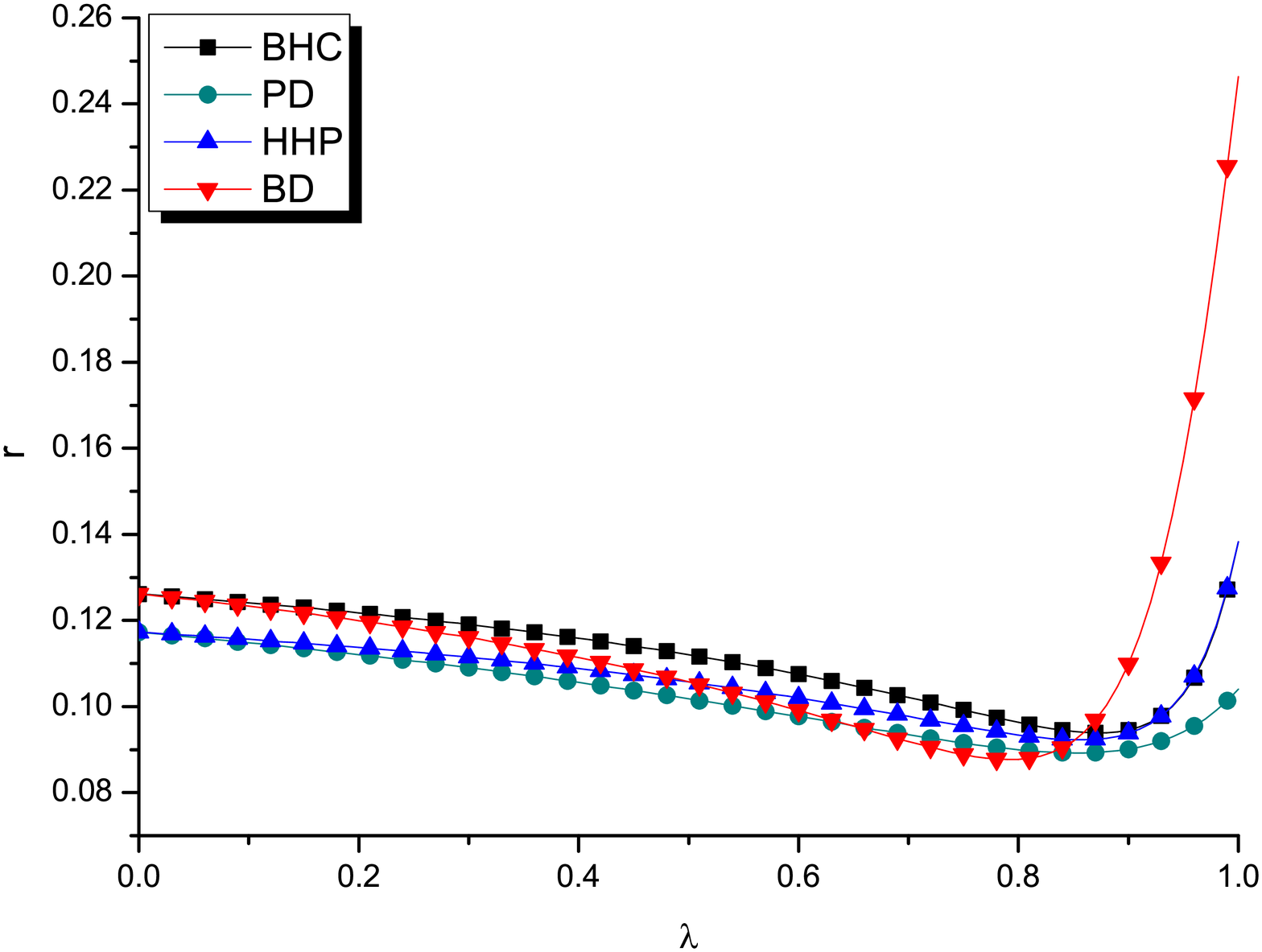}}
  \subfigure{
    \label{fig2:subfig:b} 
    \includegraphics[width=3in]{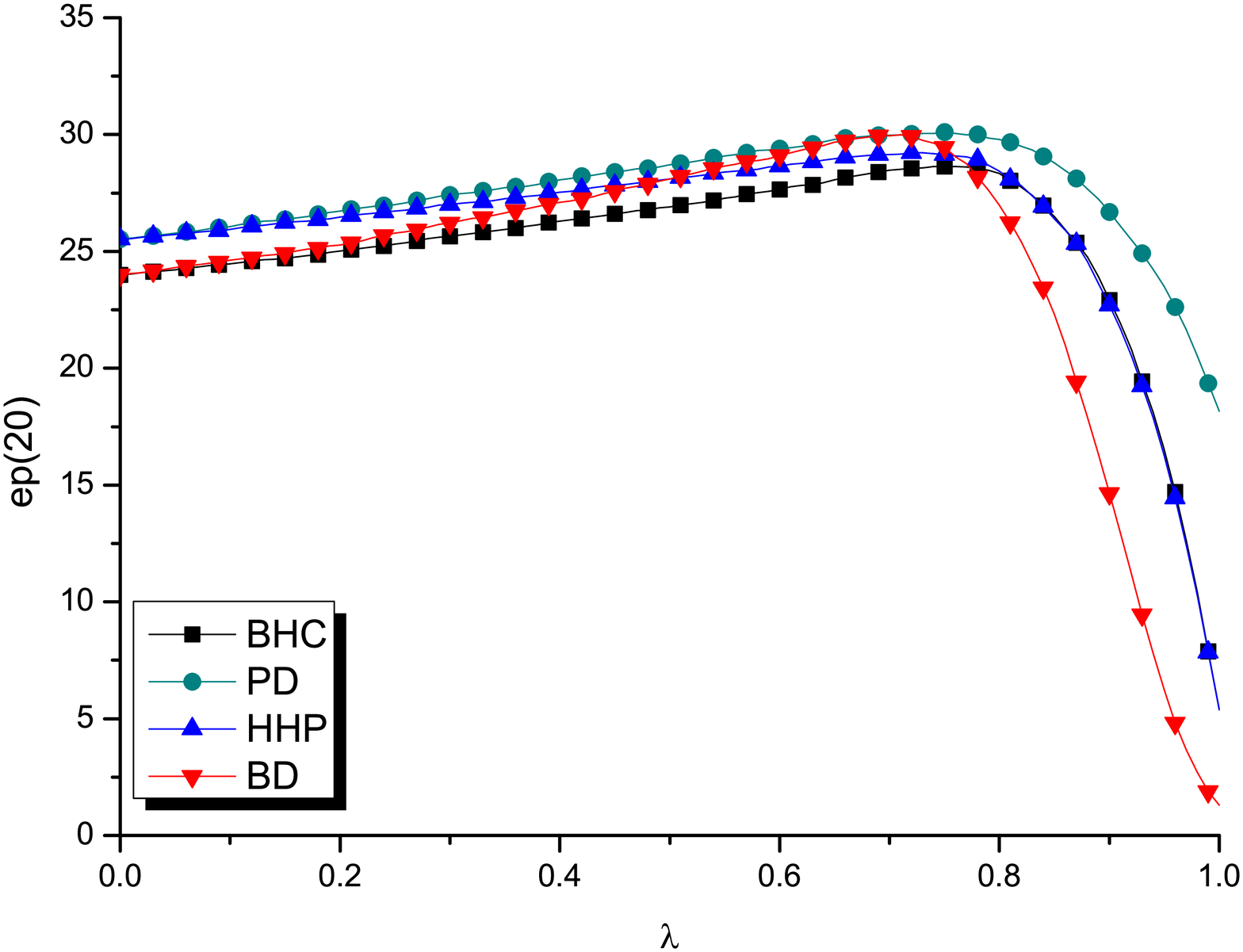}}\\
  \subfigure{
    \label{fig2:subfig:c} 
    \includegraphics[width=3in]{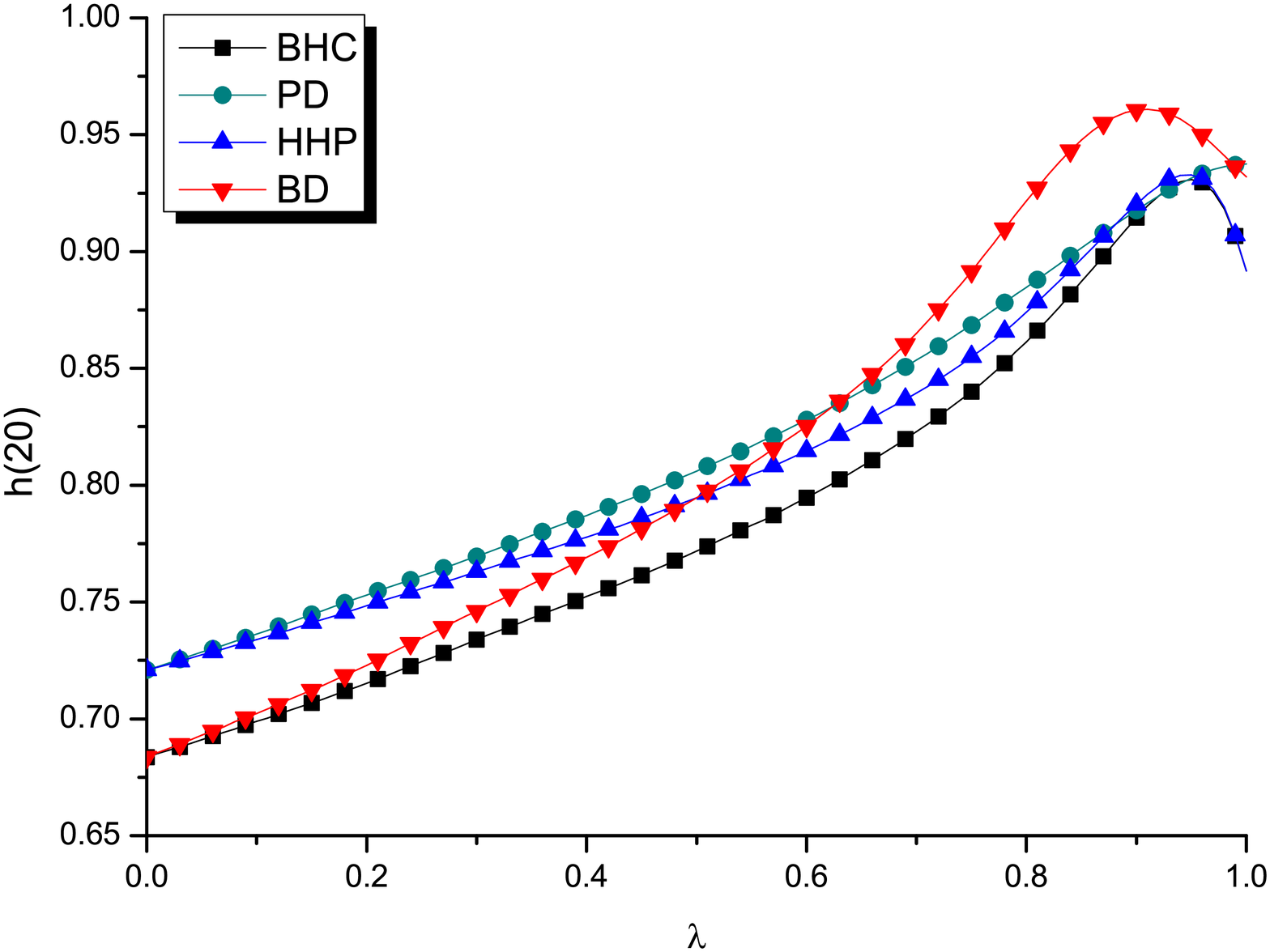}}
   \subfigure{
    \label{fig2:subfig:d} 
    \includegraphics[width=3in]{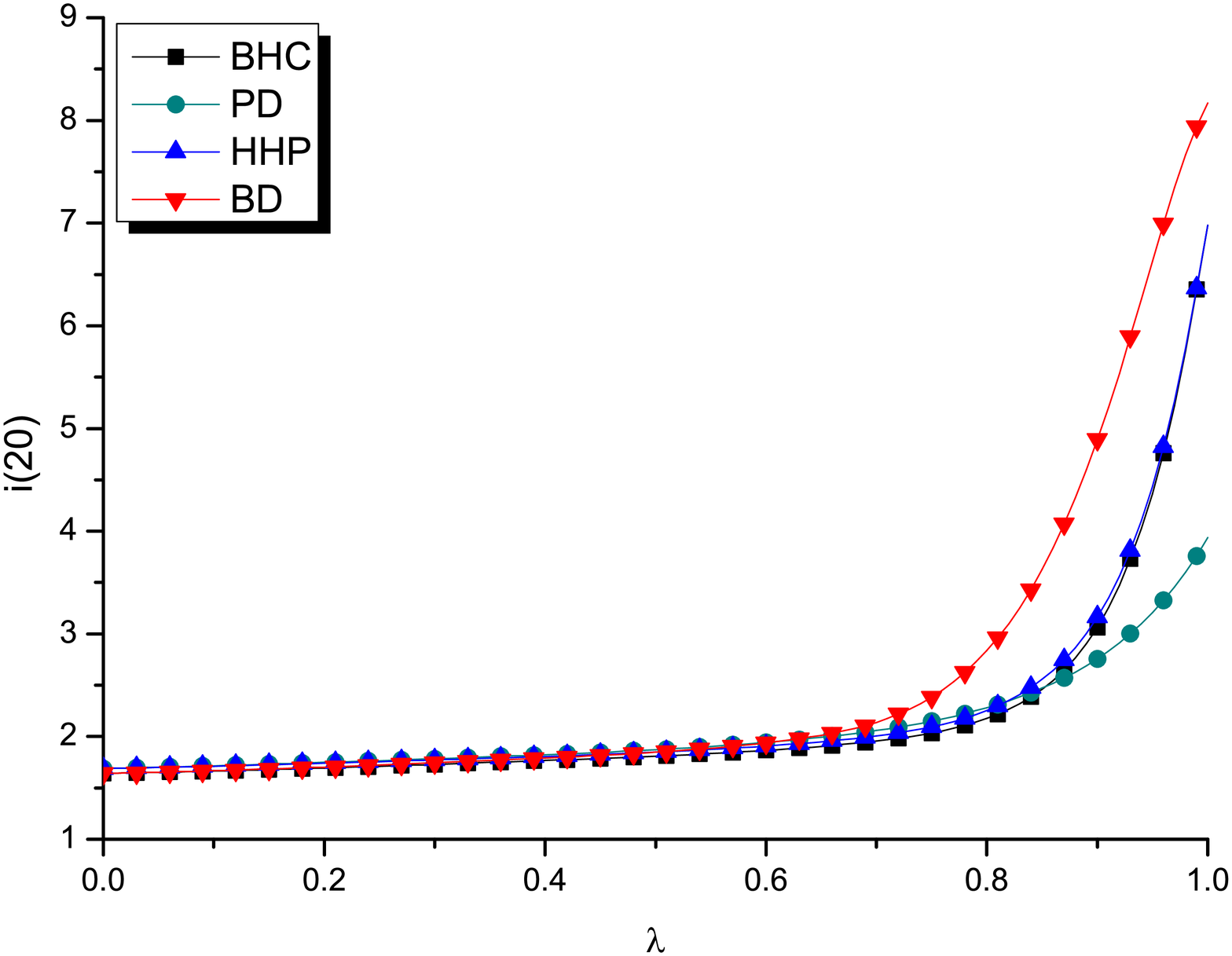}}
  \caption{(Color online) The recommendation results of four
  algorithms on Movielens data set.}
  \label{fig2} 
\end{figure}

\begin{figure}
  \centering
  \subfigure{
    \label{fig3:subfig:a} 
    \includegraphics[width=3in]{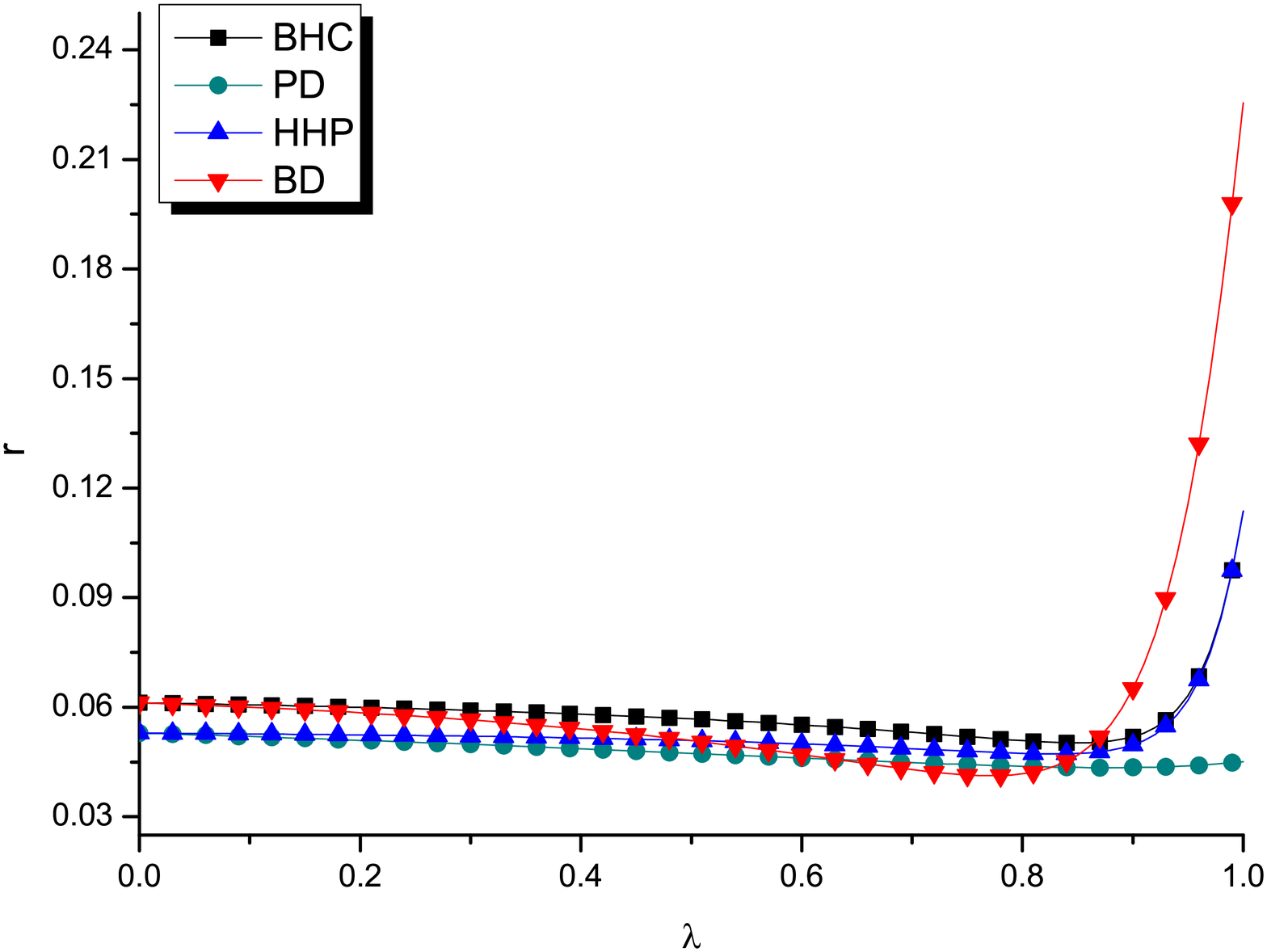}}
  \subfigure{
    \label{fig3:subfig:b} 
    \includegraphics[width=3in]{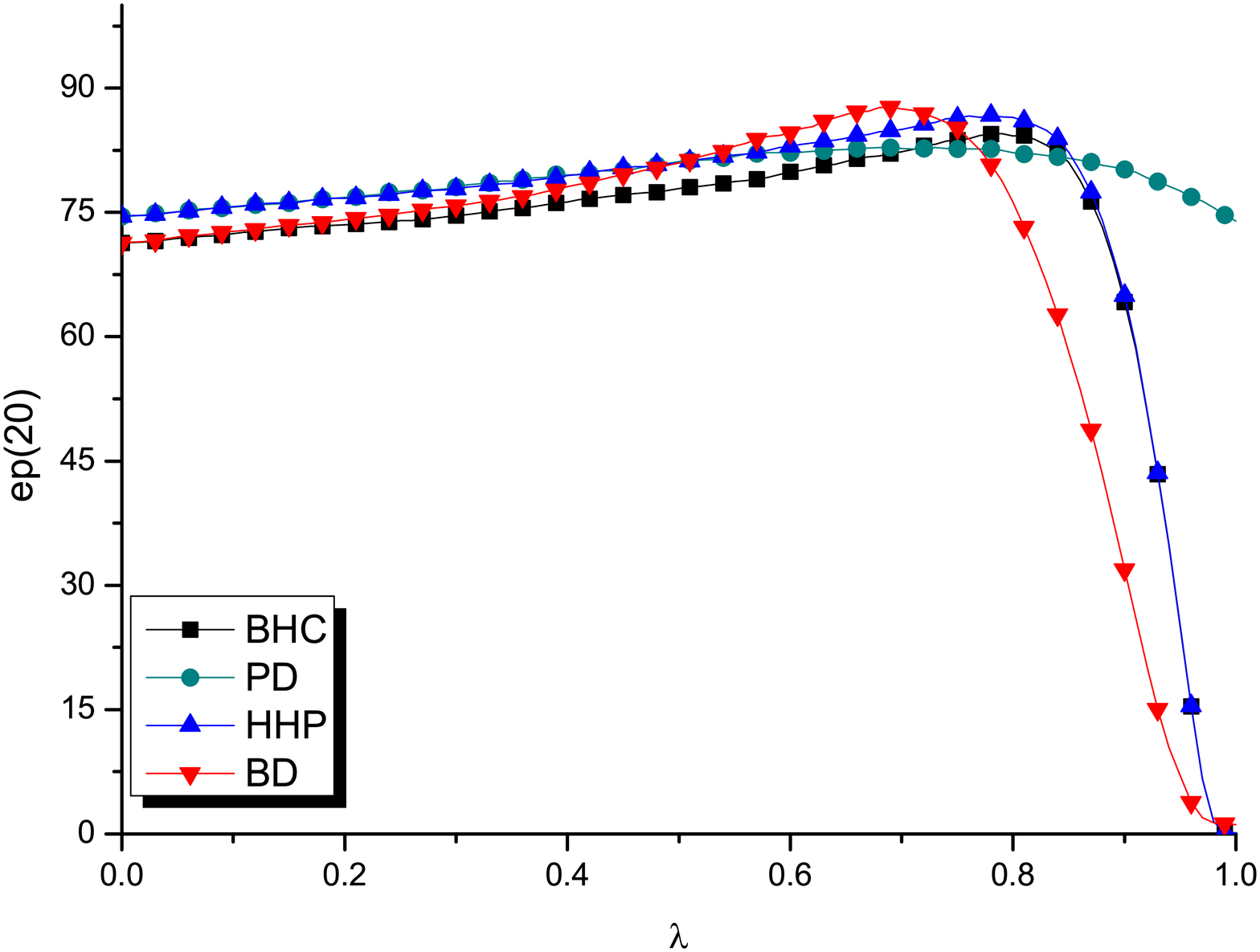}}\\
  \subfigure{
    \label{fig3:subfig:c} 
    \includegraphics[width=3in]{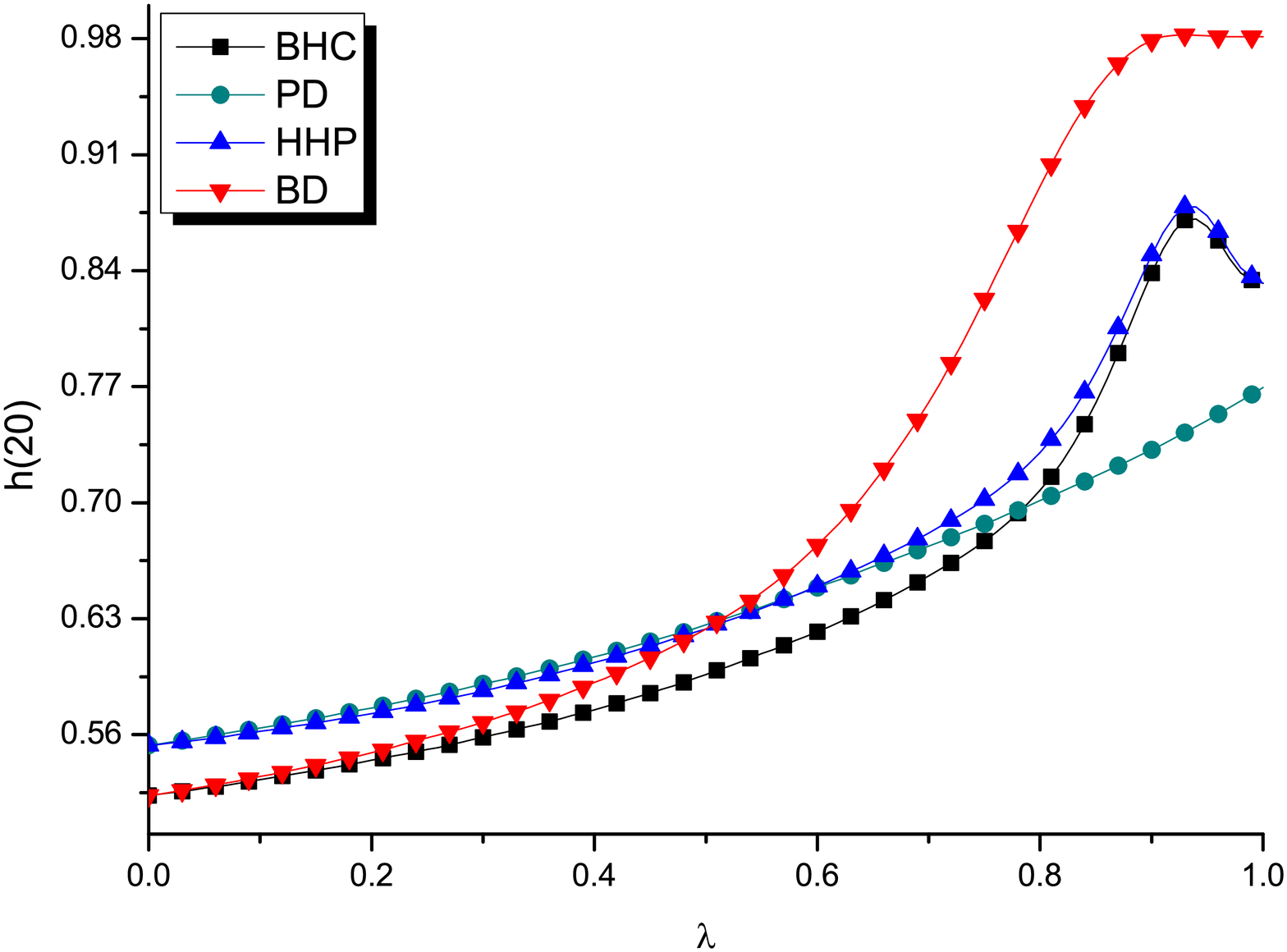}}
   \subfigure{
    \label{fig3:subfig:d} 
    \includegraphics[width=3in]{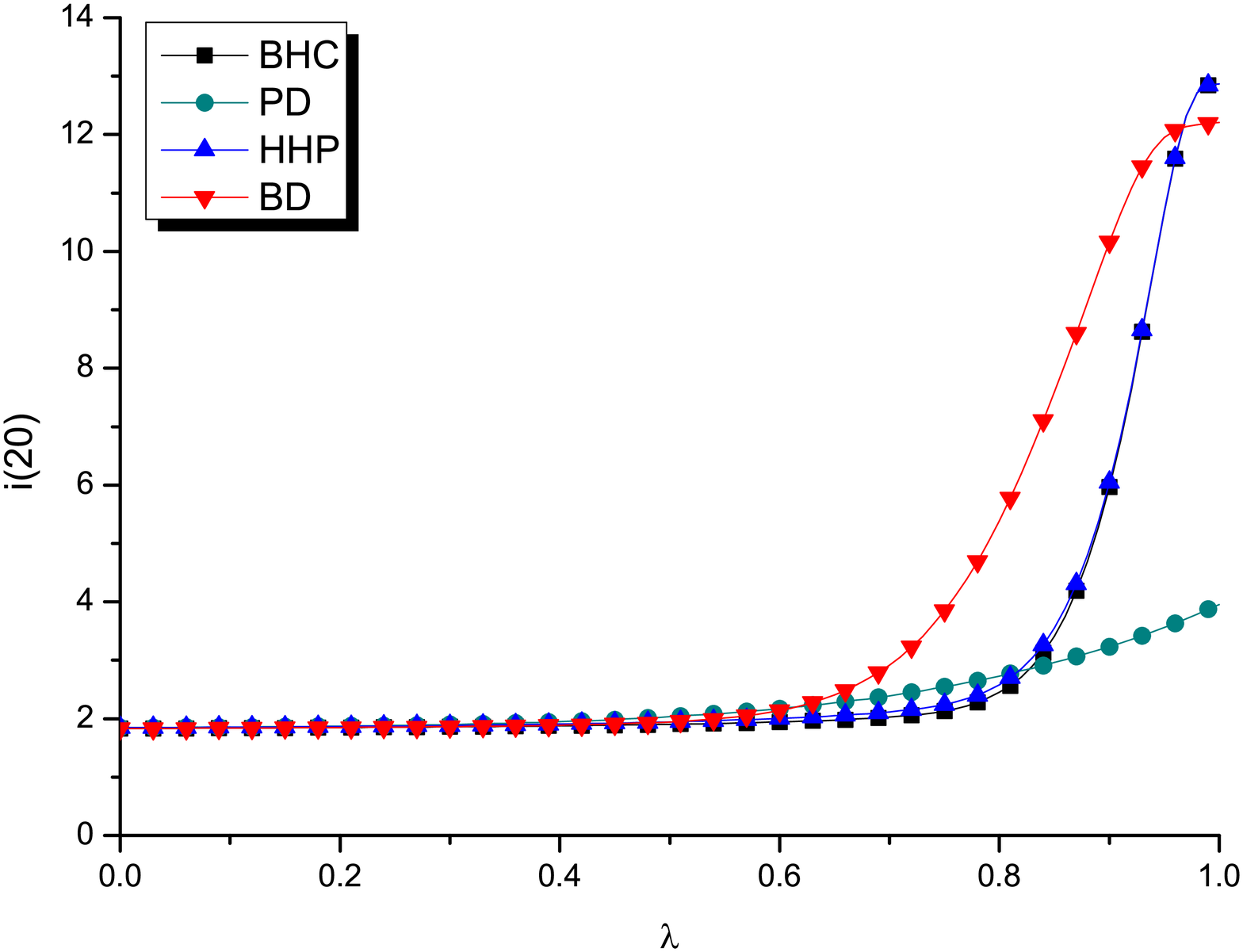}}
  \caption{(Color online) The recommendation results of four
  algorithms on Netflix data set.}
  \label{fig3} 
\end{figure}

\begin{figure}
  \centering
  \subfigure{
    \label{fig4:subfig:a} 
    \includegraphics[width=3in]{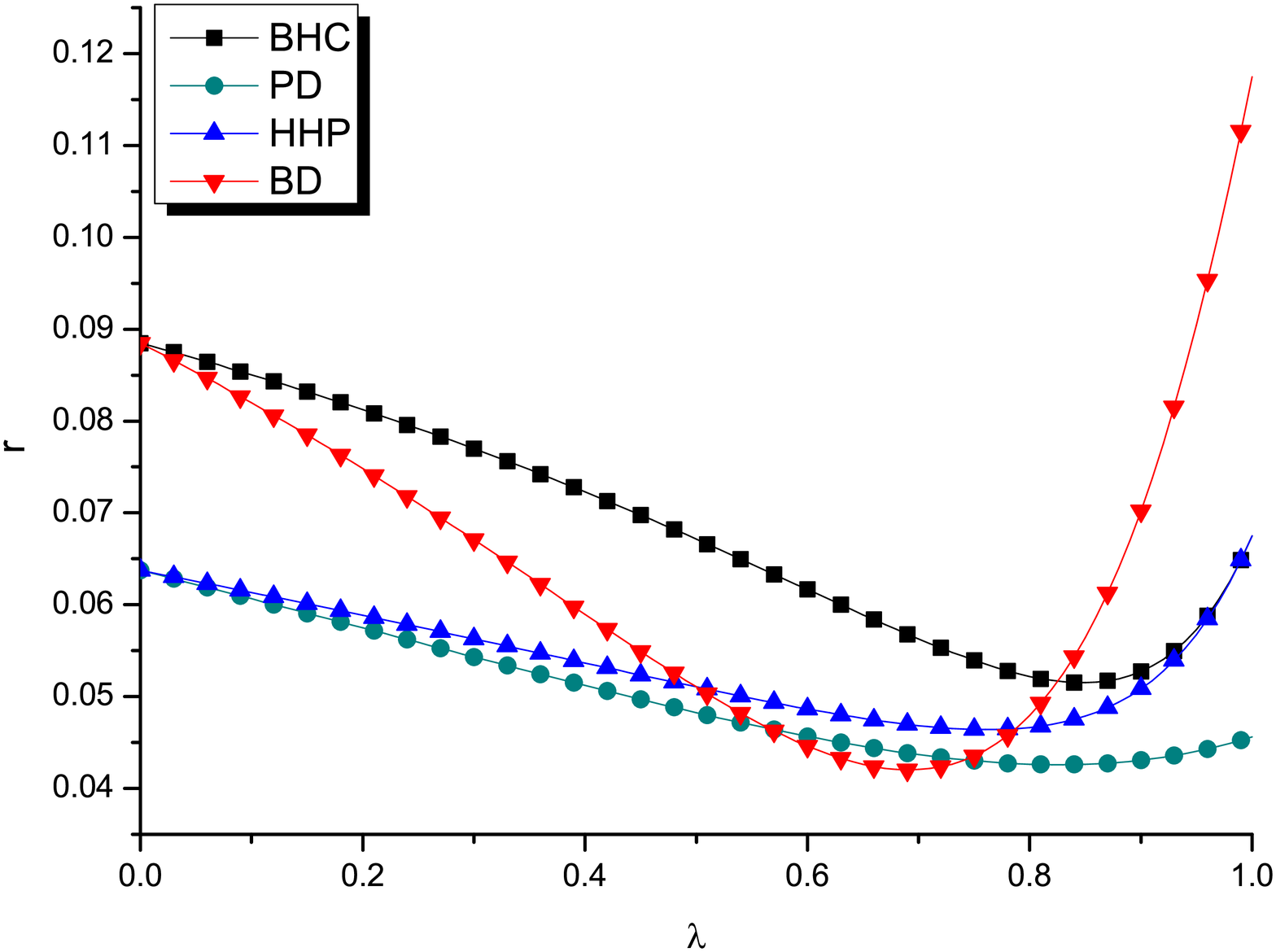}}
  \subfigure{
    \label{fig4:subfig:b} 
    \includegraphics[width=3in]{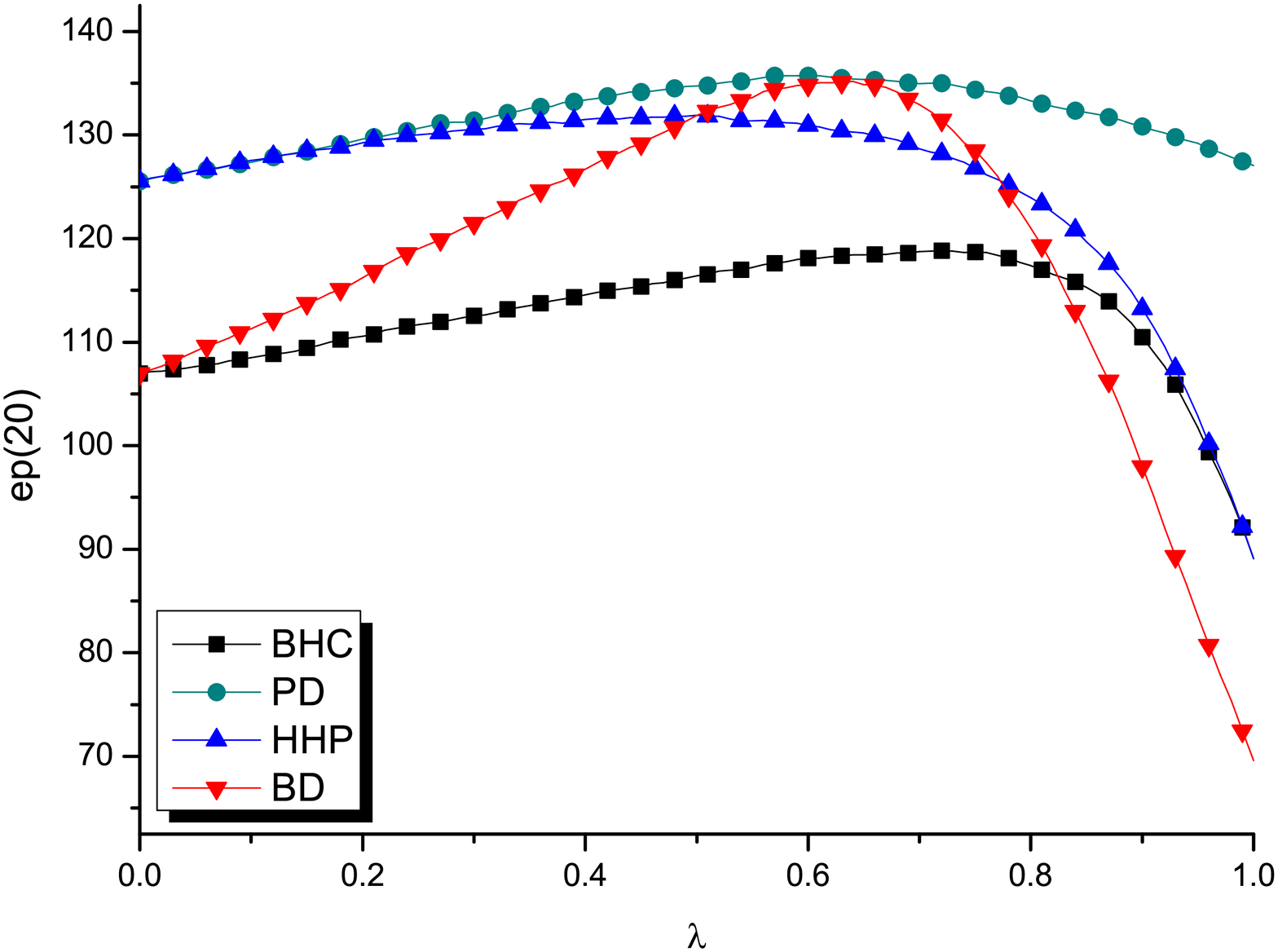}}\\
  \subfigure{
    \label{fig4:subfig:c} 
    \includegraphics[width=3in]{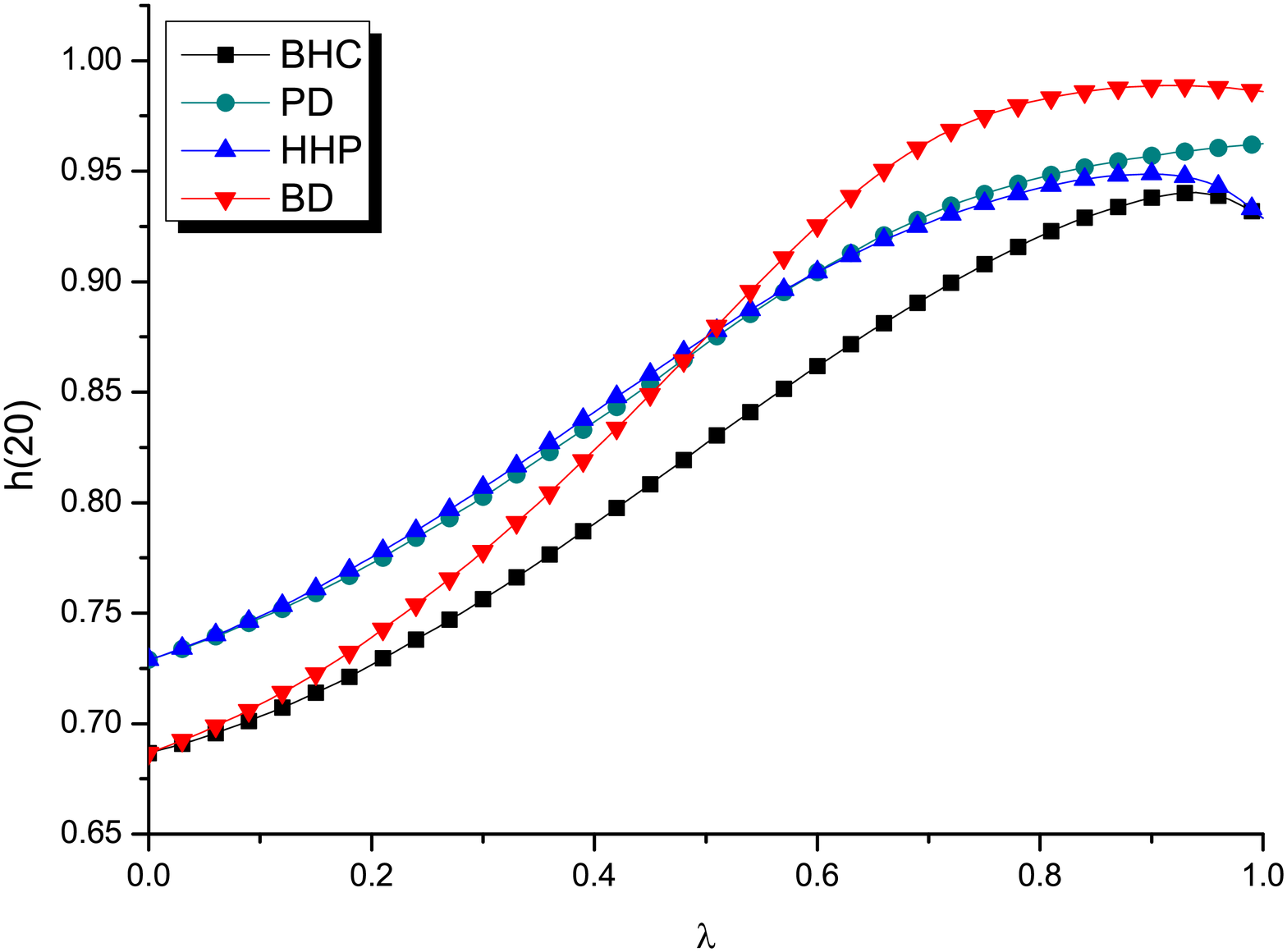}}
   \subfigure{
    \label{fig4:subfig:d} 
    \includegraphics[width=3in]{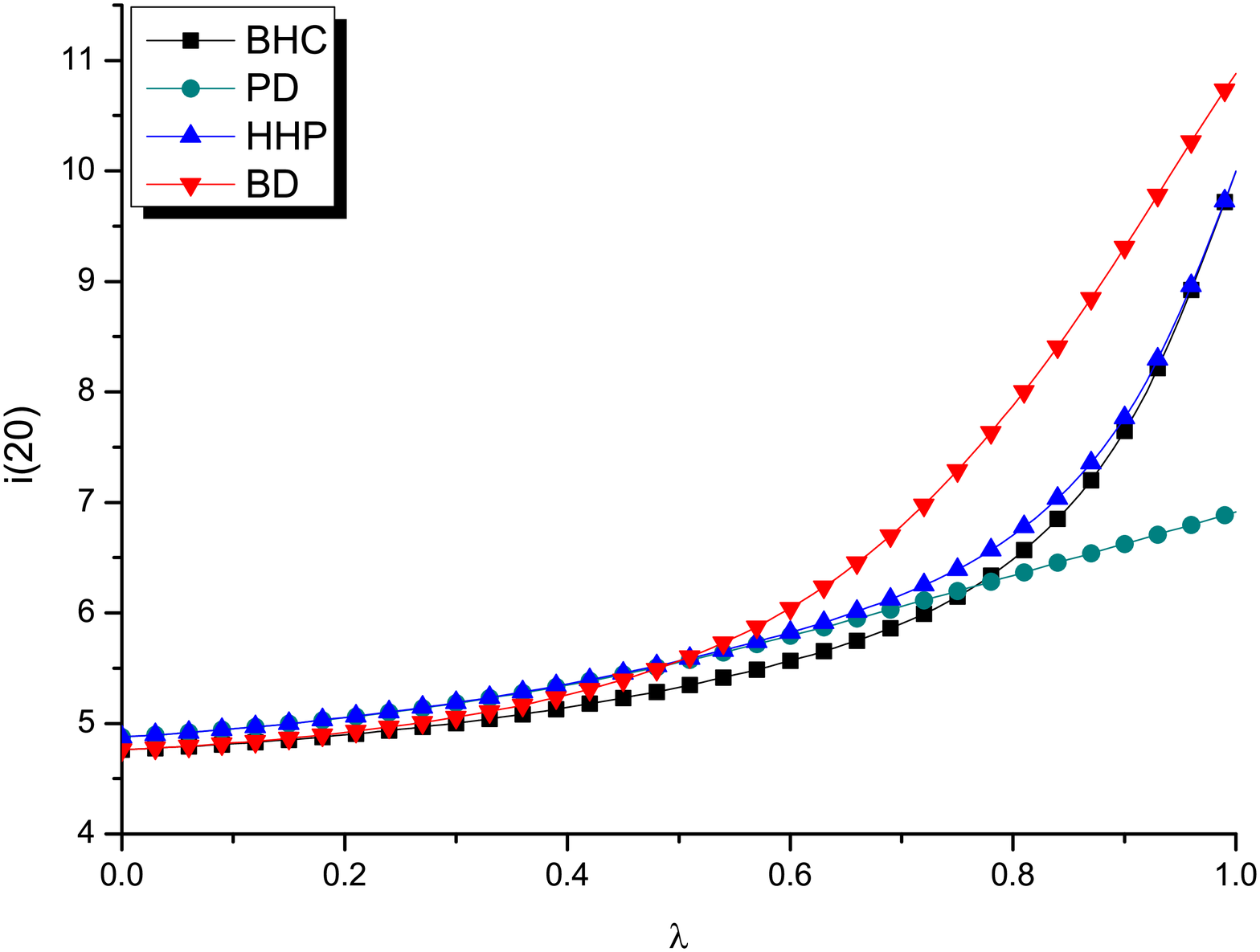}}
  \caption{(Color online) The recommendation results of four
  algorithms on RYM data set.}
  \label{fig4} 
\end{figure}

In order to show our algorithm's superior performance, we compare
BD with HHP, BHC and PD. In Fig.\ref{fig2} - Fig.\ref{fig4}, to compare those four algorithms conveniently and show the results in the same scale, we use $\lambda$ instead of $1-\lambda$ and $-\varepsilon$ for HHP and PD algorithms, respectively. Fig.\ref{fig2}, Fig.\ref{fig3} and
Fig.\ref{fig4} show the performance of the four algorithms under
different $\lambda$ on $MovieLens$, $Netflix$ and $RYM$ data sets,
respectively. Summaries of the results for the four algorithms and
metrics on $MovieLens$, $Netflix$ and $RYM$ data sets are shown
respectively in Table.\ref{table2}, Table.\ref{table3} and
Table.\ref{table4}. Clearly, BD outperforms HHP and BHC on
$MovieLens$ data set, outperforms BHC and PD on $Netflix$ data set
and outperforms HHP and PD on $RYM$ data set. Among all four
precious algorithms, BD gives the best ranking score and hamming
distance. PD and HHP give the highest precision on $MovieLens$ and
$Netflix$ data sets, respectively. BHC gives the best novelty on
$RYM$ data set. Comparing these four outstanding algorithms, we find
that BD gives almost the best accuracy and provides the much more
diversity results. For instance, in $MovieLens$, the BD algorithm
decreases the ranking score to 0.08769 while simultaneously improves
the hamming distance and novelty to 0.91572 and 2.7269,
respectively. Meanwhile, the BD algorithm's precision is 27.63,
which is near the best value by PD.

\begin{figure}
  \centering
  \subfigure{
    \label{fig5:subfig:a} 
    \includegraphics[width=2in]{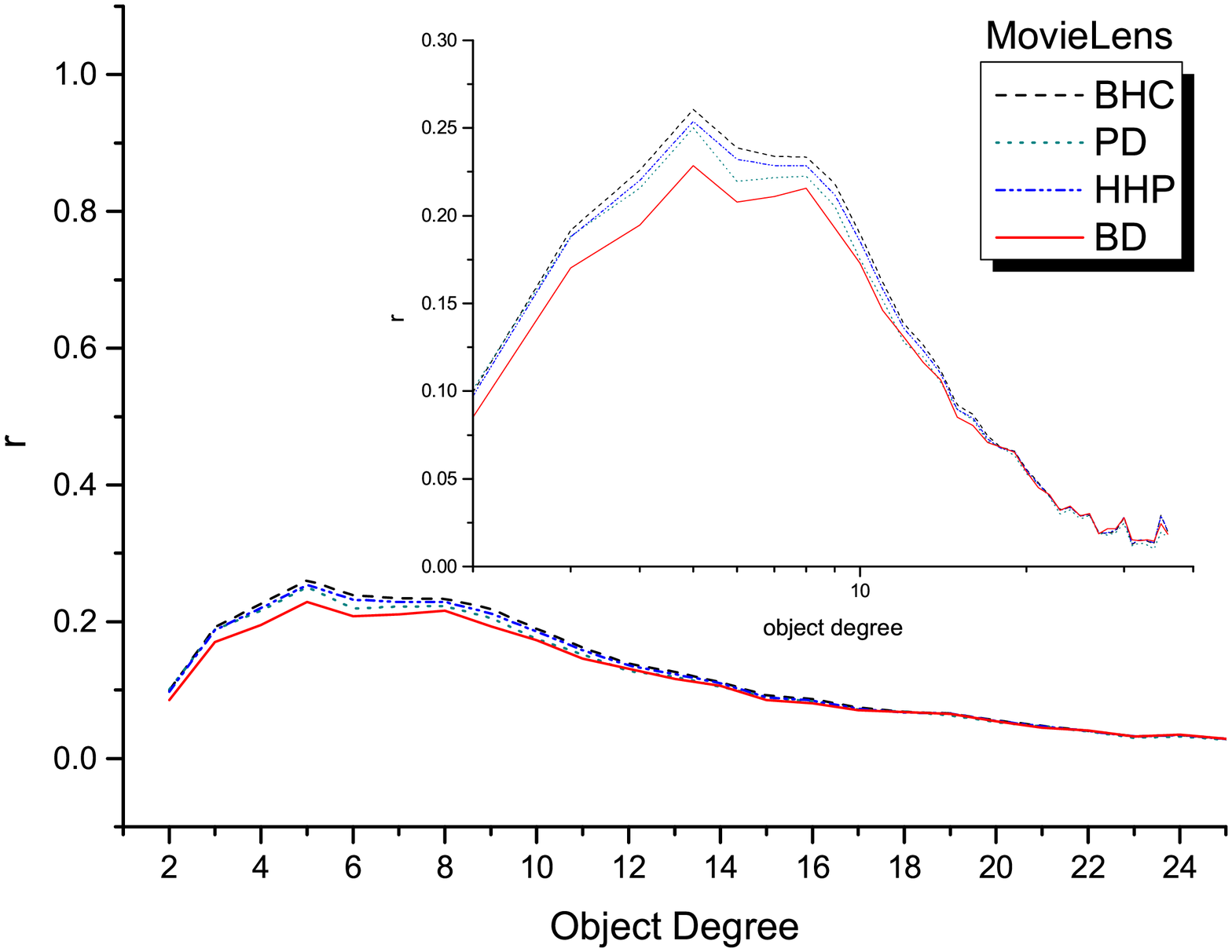}}
  \subfigure{
    \label{fig5:subfig:b} 
    \includegraphics[width=2in]{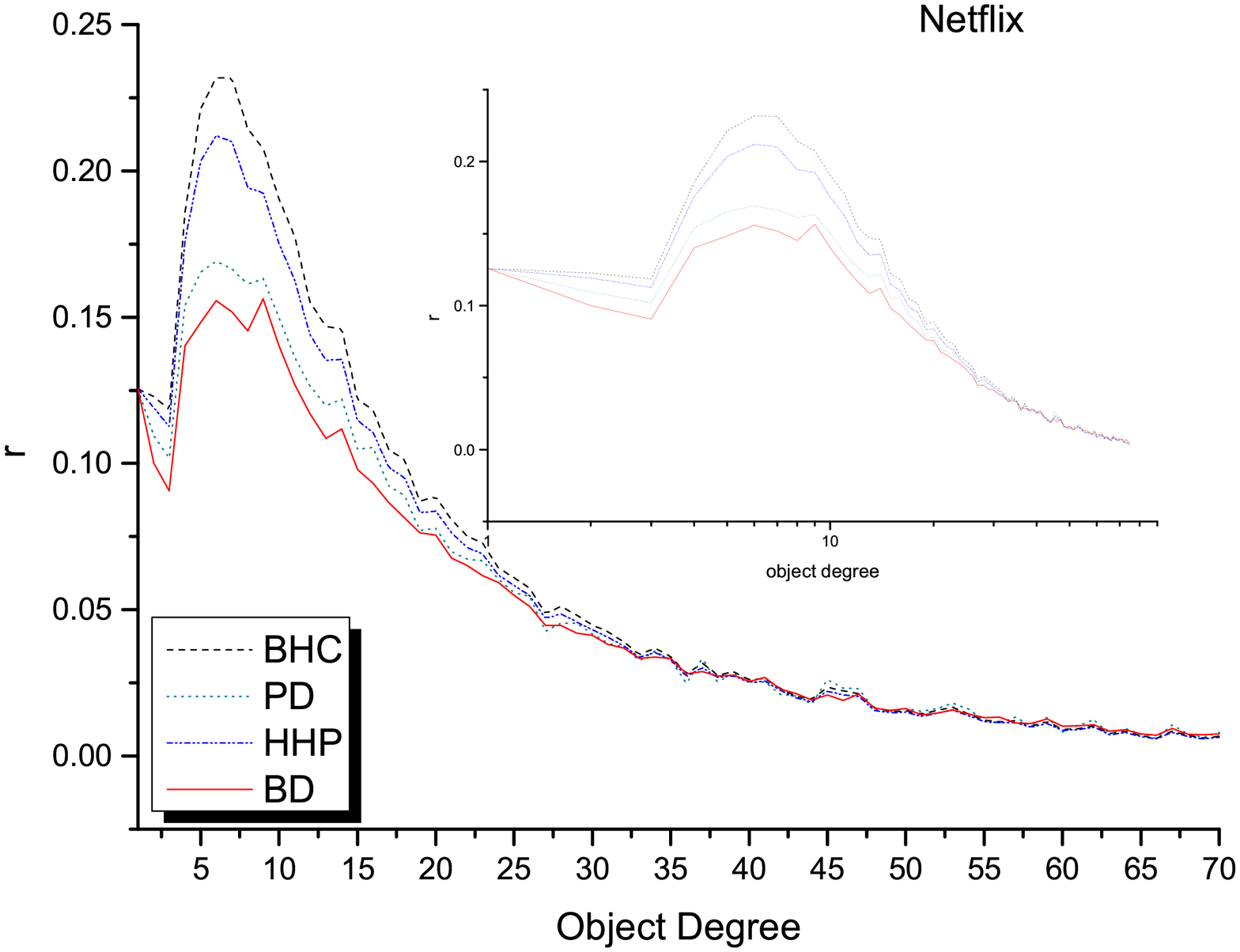}}
  \subfigure{
    \label{fig5:subfig:b} 
    \includegraphics[width=2in]{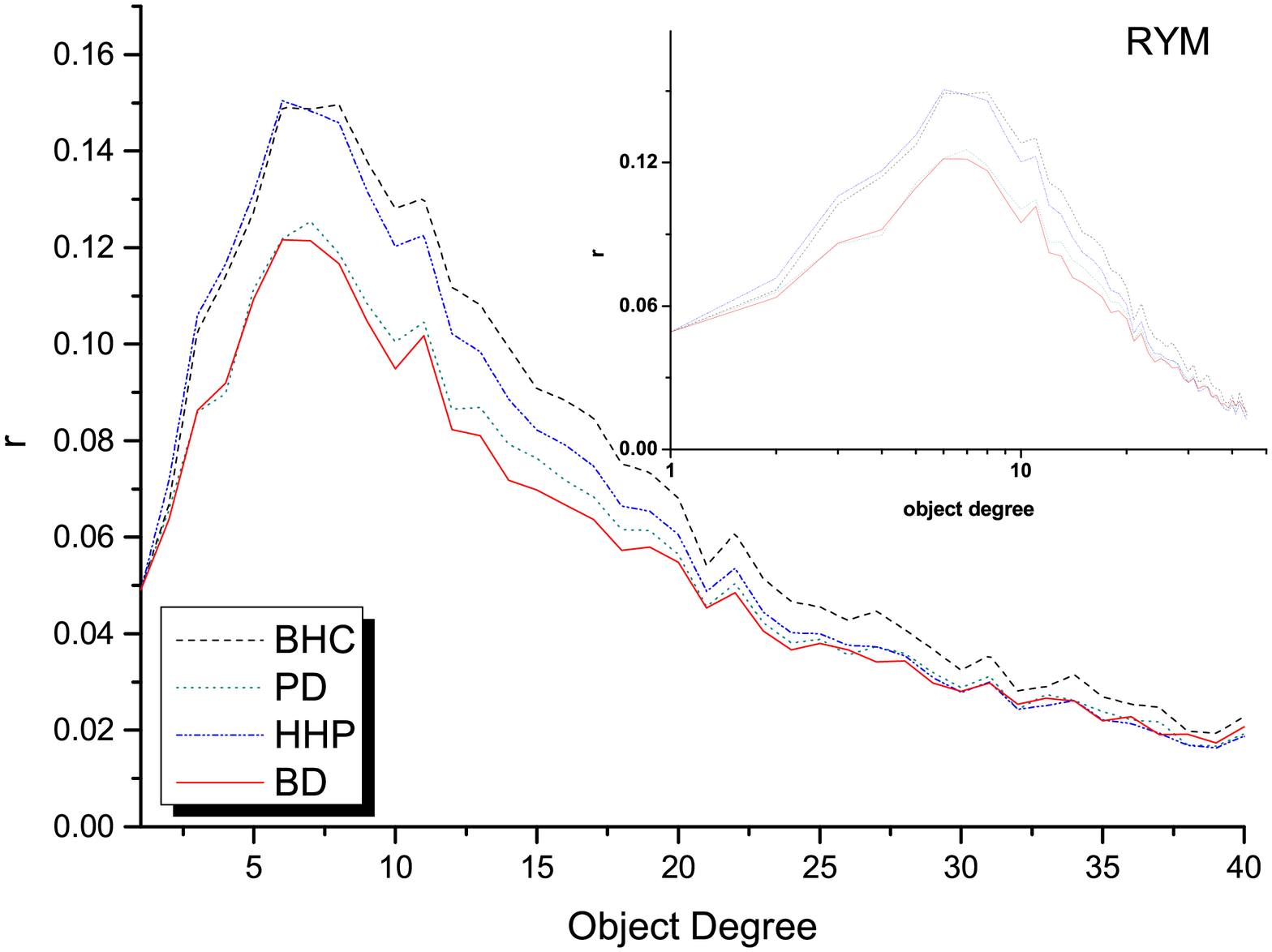}}
  \caption{{\bf (Color online) Dependence of ranking score $r$ on the object degree.}
  For a given $x$, its corresponding $r$ is obtained by averaging over
all objects whose degrees are in the range of $[a(x^2-x),a(x^2+x)]$,
where $a$ is chosen as $\frac{1}{2}$log5 for a better illustration.
The inset figure amplifies that $r$ versus the degree of objects.}
  \label{fig5} 
\end{figure}

In Fig.\ref{fig5}, we investigate the dependence of ranking  score
on the object degree on $MovieLens$, $Netflix$ and $RYM$ data sets.
For a given $x$, its corresponding $r$ is obtained by averaging over
all objects whose degrees are in the range of $[a(x^2-x),a(x^2+x)]$,
where $a$ is chosen as $\frac{1}{2}$log5 for a better illustration.
The inset figure amplifies that $r$ versus the degree of objects.
Generally speaking, on average, the popular objects have more
opportunity to be recommended than unpopular objects and can be more
accurately recommended. The inset figures show the difference of
these four algorithms' ability to accurately recommend the unpopular
objects. Clearly, our algorithm BD has the best ability for this
end which is followed by PD. Moreover, by comparing BD with PD and
HHP, we find that although they both consider the mass diffusion and
heat conduction processes, giving the same strength at the first and
last step will have better effects on the unpopular objects.


\section*{Discussion}
To summary, we proposed a novel method to combine the mass diffusion
and heat conduction which address the accuracy-diversity dilemma in
recommender systems. We found that the smallest ranking score values
are almost unchanged when giving the same strength on mass and heat
diffusion processes. Therefore, we use only one parameter to achieve
the optimal performance. Numerical results on three benchmark data
indicate that the accuracy and diversity are simultaneously improved
on our algorithm. In addition, we compare our algorithm with other
three outstanding algorithms and find out that our algorithm has the
best performance. Moreover, we investigate the four algorithms'
ability to accurately recommend the unpopular objects, and found our
algorithm BD has the best ability on it.

In a real online recommender system, generally speaking, most  of
the large-degree objects are popular objects, which have large
weights and easily to be recommended while the small-degree objects
might be difficult to be recommended. Therefore, the ability to
accurately recommend the unpopular objects is an important issue in
recommender systems. In a word, this work provides a practical
solution for online recommendation on how to promote the attention
of the long-tailed products.

This article only provides a simple method to combine the mass
diffusion and heat conduction by giving them the same weights, while
a couple of issues remain open for future study. First, we are lack
of quantitative understanding of the structure and dynamics of
information network. Second, the impact of social network is
overlooked in recommendation systems, although the relation between
social influence and recommender systems is not clear thus far, we
deem that an in-depth understanding of social network should be
helpful for better recommendations. Finally, the multi-layered
network consists of social network and information network can be
taken into account to describe the underlying hierarchical
structure, thus the Social Network Analysis (SNA) based techniques
can be used to provide more substantial recommendations, and social
predictions as well.

\section*{Materials and Methods}

\subsection*{Dataset Description}

To test our algorithm's performance, we employ three different
datasets (see Table.\ref{tab:1} for basic statistics). The
$MovieLens$ data set was collected by the GroupLens Research Project
at the University of Minnesota. The data was collected through the
MovieLens web site (movielens.umn.edu) during the seven-month period
from September 19, 1997 through April 22, 1998. The $Netflix$
dataset is a randomly selected subset of the huge dataset provided
for the Netflix Prize \cite{bennett2007netflix}. The $RYM$ data set is obtained by downloading publicly available data
from the music rating website RateYourMusic.com. We use the
information of the links between users and objects.

\subsection*{Metrics}

Accuracy is the most important measure in evaluating the performance
of recommendation algorithms. A good algorithm is expected to give
accurate recommendations, namely higher ability to find what users
like. In order to measure the recommendation accuracy, we make use
of ranking score $r$ \cite{lu2011information} and precision
enhancement $ep(L)$ \cite{zhou2010solving}. For a target user, the
recommender system will return a ranking list of all uncollected
objects to him/her. For each link in test set, we measure the
rank($r_{i \alpha}$) of this object in the recommendation list of
this user.
\begin{center}
\begin{equation}
\label{equ8} r_{i\alpha} = \frac{p_{\alpha}}{l_i}
\end{equation}
\end{center}
where $p_{\alpha}$ means that object $\alpha$ is listed in the place
$p_{\alpha}$ of the ranking list of user $i$, $l_i$ is the number of
links of user $i$ in the test set.

\begin{center}
\begin{equation}
\label{equ9} r = \frac{1}{|E_P|}\sum_{i\alpha \in E_P}{r_{i\alpha}}
\end{equation}
\end{center}
where $i\alpha$ denotes the link connecting $u_i$  and $o_{\alpha}$
in the test set.

A random recommendation will randomly choose $L$ objects  from the
training data for a target user, so we consider the precision
values($ep(L)$) relative to the precision of random recommendations.

\begin{center}
\begin{equation}
\label{equ10} ep(L) =
\frac{1}{m}\sum_{l=1}^{m}\frac{n}{L}\frac{p}{k_{l_{test}}}
\end{equation}
\end{center}
where $p$ is the number of objects in the data sets  and
$k_{l_{test}}$ is the degree of user $l$ in test data.

Beside accuracy, diversity is taken into account as another
important metric to evaluate the recommendation algorithm. In order
to measure the recommendation diversity and novelty, we make use of
Hamming distance ($h(L)$ for short) \cite{zhou2008effect} and self
information $I(L)$ in \cite{zhou2010solving}, respectively.

\begin{center}
\begin{equation}
\label{equ11} h_{ij}(L) = 1 - \frac{q_{ij}(L)}{L}
\end{equation}
\end{center}
where $q_{ij}$ is the number of common objects in  the top L places
of both lists of user $i$ and user $j$. Averaging $h_{ij}(L)$ over
all pairs of users existing in the test set, we obtain the mean
distance $h(L)$, for which greater value means greater
personalization of user's recommendation lists.

The $I(L)$ concerns the capacity of the recommender  system to
generate novel and unexpected results. Given an object $\alpha$, the
chance a randomly selected user has collected it is $k_{\alpha}/u$,
thus its self-information is $I_{\alpha} = log_2(u/k_{\alpha})$, so
$I(L)$ is:
\begin{center}
\begin{equation}
\label{equ12} I(L) = \sum_{l=1}^{m}\sum_{\alpha=1}^{L}{I_{\alpha}}
\end{equation}
\end{center}
where $L$ is the length of the recommendation list.

\subsection*{Baseline Algorithms}
The original recommendation algorithm mimicking mass diffusion (MD)
process is referred to as Network-Based Inference (NBI) \cite{zhou2007bipartite}, and ProbS \cite{zhou2010solving},
respectively. The initial resource vector $\textbf{f}$ is defined as
$f_{\alpha} = a_{l\alpha}$ where $a_{l\alpha}=1$ if user $l$ has
collected object $\alpha$, otherwise $a_{l\alpha}=0$. The element
$w_{\alpha\beta}$ of the transfer matrix $W$ is written as

\begin{center}
\begin{equation}
\label{equ1} w_{\alpha \beta} =
\frac{1}{k_{o_{\beta}}}\sum_{l=1}^{m}{\frac{a_{l\alpha}a_{l\beta}}{k_{u_l}}},
\end{equation}
\end{center}
where $k_{o_{\beta}} = \sum_{i = 1}^{m}{a_{i\beta}}$ and $k_{u_l} =
\sum_{r = 1}^{n}{a_{lr}}$ denote the degrees of object $o_{\beta}$
and user $u_l$, respectively.

The algorithm analogous to heat conduction (HC) is called HeatS in
\cite{zhou2010solving}. The significant difference between HC and
MD is that HC redistributes a resource via a nearest-neighbor
averaging process, while MD works by equally distributing the
resource to the nearest neighbor. The initial $\textbf{f}$ of HC
is the same with MD. The difference lies in the transfer matrix:
\begin{center}
\begin{equation}
\label{equ2} w_{\alpha \beta} = \frac{1}{k_{o_{\alpha}}}
\sum_{l=1}^{m}{\frac{a_{l\alpha}a_{l\beta}}{k_{u_l}}},
\end{equation}
\end{center}

As we know, MD has high recommendation accuracy yet low
diversity, while HC, which is designed specifically to address
the challenge of diversity, has relatively low accuracy. Many
researchers attempted to solve this diversity-accuracy dilemma and
have found some effective ways:

In \cite{zhou2010solving}, the authors proposed a nonlinear hybrid
method to combine MD and HC, called HHP algorithm, by
introducing a hybridization parameter $\lambda$ into the transfer
matrix $W$:
\begin{center}
\begin{equation}
\label{equ3} w_{\alpha \beta} =
\frac{1}{{k_{o_{\alpha}}}^{1-\lambda} {k_{o_{\beta}}}^{\lambda}}
\sum_{l=1}^{m}{\frac{a_{l\alpha }a_{l\beta}}{k_{u_l}}},
\end{equation}
\end{center}
where $\lambda = 0$ gives the pure HeatS and $\lambda = 1$ gives the
pure MD, which makes a trade-off between diversity and accuracy
by adjusting the tunable parameter $\lambda$.

Motivated by enhancing the ability to find unpopular and niche
objects, L\"{u} et al. \cite{lu2011information} proposed a
preferential diffusion method (PD for short) based on a hybrid of
MD and HC with preference on MD. By changing the amount of
resource that an object $o_{\alpha}$ receives in the last step to
$k_{o_{\alpha}}^{\varepsilon}$ where $-1 \leq \varepsilon \leq 0$ is
a free parameter, the resource transfer matrix reads:
\begin{center}
\begin{equation}
\label{equ4} w_{\alpha \beta} = \frac{1}{{k_{o_{\beta}}}
{k_{o_{\alpha}}^{-\varepsilon}}} \sum_{l=1}^{m}{\frac{a_{l\alpha
}a_{l\beta}}{\mathcal{M}}},
\end{equation}
\end{center}
where
$\mathcal{M}=\sum_{r=1}^{n}{{a_{lr}{k_{o_\alpha}^{\varepsilon}}}}$.
Clearly, when $\varepsilon = 0$, it gives us the pure MD
algorithm.

In \cite{liu2011information}, the authors proposed a Biased Heat
Conduction (BHC for short) method based on HC. By decreasing the
temperatures of small-degree objects, BHC could simultaneously
enhance the accuracy and diversity. The element $w_{\alpha\beta}$ of
the transfer matrix $W$ is:
\begin{center}
\begin{equation}
\label{equ5} w_{\alpha \beta} = \frac{1}{k_{o_{\alpha}}^{\lambda}}
\sum_{l=1}^{m}{\frac{a_{l\alpha }a_{l\beta}}{k_{u_l}}},
\end{equation}
\end{center}
where $0\leq \lambda \leq 1$, which indicates that the influence of
large-degree objects would be strengthened in the last diffusion
step.

\section*{Acknowledgments}
We acknowledge Zi-Ke Zhang, Jun-Lin Zhou and Xu-Zhen Zhu for helpful discussions and irradiative ideas. This work was partially supported by Natural Science Foundation of
China (Grant Nos. 61103109, 11105024 and 61300018), Special Project
of Sichuan Youth Science and Technology Innovation Research Team
(Grant No. 2013TD0006).

\bibliography{references}

\section*{Figure Legends}

%
%


\section*{Tables}
\begin{table}[!ht]
\centering
\caption{
\bf{Properties of the tested datasets}}
\begin{tabular}{ccccc}
\hline
Data sets & Users & Objects & Links & Sparsity\\
\hline
MovieLens  & 943 & 1,682 & 100,000  & $6.30\times{10^{-2}}$ \\
Netflix  & 10,000 & 5,640 & 701,947 & $1.24\times{10^{-2}}$ \\
RYM & 33,762 &5,267 & 675,817 & $3.8\times{10^{-3}}$\\
\hline
\end{tabular}
\begin{flushleft}
\end{flushleft}
\label{tab:1}
\end{table}

\begin{table}[!ht]
\centering \caption{ {\bf Algorithmic performance for MovieLens data
set.} The optimal parameters are $\lambda_{opt}=0.14$ for HHP,
$\lambda_{opt}=0.87$ for BHC, $\varepsilon_{opt}=-0.85$ for PD and
$\lambda_{opt}=0.79$ for BD.}
\begin{tabular}{ccccc}
\hline
Method & $r$ & $ep(20)$ & $h(20)$ & $I(20)$\\
\hline
HHP  & 0.09228 & 25.892 & 0.90162  & 2.6452 \\
BHC  & 0.09388 & 25.367 & 0.89809 & 2.6474 \\
PD & 0.08924 &\textbf{28.793} & 0.90146 & 2.4716\\
BD&\textbf{0.08769}&27.63&\textbf{0.91572}&\textbf{2.7269}\\
\hline
\end{tabular}
\begin{flushleft}
\end{flushleft}
\label{table2}
\end{table}

\begin{table}[!ht]
\centering \caption{ {\bf Algorithmic performance for Netflix data
set.} The optimal parameters are $\lambda_{opt}=0.17$ for HHP,
$\lambda_{opt}=0.85$ for BHC, $\varepsilon_{opt}=-0.88$ for PD and
$\lambda_{opt}=0.77$ for BD.}
\begin{tabular}{ccccc}
\hline
Method & $r$ & $ep(20)$ & $h(20)$ & $I(20)$\\
\hline
HHP  & 0.04719 & \textbf{84.89511} & 0.75589  & 3.03893 \\
BHC  & 0.05023 & 81.02979 & 0.76022 & 3.41337 \\
PD & 0.04348 &81.02979 & 0.76022 & 3.41337\\
BD&\textbf{0.04125}&82.53697&\textbf{0.85025}&\textbf{4.38952}\\
\hline
\end{tabular}
\begin{flushleft}
\end{flushleft}
\label{table3}
\end{table}

\begin{table}[!ht]
\centering \caption{ {\bf Algorithmic performance for RYM data set.}
The optimal parameters are $\lambda_{opt}=0.24$ for HHP,
$\lambda_{opt}=0.85$ for BHC, $\varepsilon_{opt}=-0.82$ for PD and
$\lambda_{opt}=0.69$ for BD.}
\begin{tabular}{ccccc}
\hline
Method & $r$ & $ep(20)$ & $h(20)$ & $I(20)$\\
\hline
HHP  & 0.04642 & 126.15419 & 0.93689  & 6.44823 \\
BHC  & 0.05151 & 115.28652 & 0.93062 & \textbf{6.95915} \\
PD & 0.04259 &122.52548 & 0.94475 & 6.85662\\
BD&\textbf{0.04202}&\textbf{133.46272}&\textbf{0.9605}&6.6999\\
\hline
\end{tabular}
\begin{flushleft}
\end{flushleft}
\label{table4}
\end{table}

\end{document}